\documentclass[runningheads,a4paper]{llncs}

\usepackage{amssymb}
\usepackage{graphicx}

\usepackage{url}
\urldef{\mailsa}\path|{karel.devogeleer,gerard.memmi}@telecom-paristech.fr,|
\urldef{\mailsb}\path|{pierre.jouvelot,fabien.coelho}@mines-paristech.fr|   
\newcommand{\keywords}[1]{\par\addvspace\baselineskip
\noindent\keywordname\enspace\ignorespaces#1}

\usepackage{stfloats}
\usepackage{array}
\usepackage[tight,footnotesize]{subfigure}

\usepackage{listings}
\usepackage{verbatim}
\usepackage{fancyvrb}

\usepackage{tikz}
\usepackage{multirow}
\usepackage{rotating}

\usepackage{mdwmath}
\usepackage{mdwtab}

\usepackage{amsmath}

\usepackage[printonlyused,smaller]{acronym}

\newcommand{\Vcc}{V_{\mathrm{CC}}}

\newcommand{\Ps}{P_{\mathrm{system}}}
\newcommand{\Pcpu}{P_{\mathrm{CPU}}}
\newcommand{\Ecpu}{E_{\mathrm{CPU}}}

\newcommand{\El}{E_{\mathrm{leak}}}
\newcommand{\Ed}{E_{\mathrm{dynamic}}}

\newcommand{\Pd}{P_{\mathrm{dynamic}}}
\newcommand{\Pt}{P_{\mathrm{total}}}
\newcommand{\Pl}{P_{\mathrm{leak}}}
\newcommand{\Psc}{P_{\mathrm{short}}}
\newcommand{\Pc}{P_{\mathrm{charge}}}

\newcommand{\Pb}{P_{\mathrm{bench}}}
\newcommand{\tb}{t_{\mathrm{bench}}}

\newcommand{\Il}{I_{\mathrm{leak}}}

\newcommand{\cb}{cc_{\mathrm{b}}}
\newcommand{\ck}{cc_{\mathrm{k}}}

\newcommand{\fo}{f_{\mathrm{opt}}}
\newcommand{\Ecpum}{E_{\mathrm{CPU,min}}}

\newcommand{\ea}{{et al.}~}

\newcommand{\BSIM}{{\small BSIM}~}

\acrodef{AAC}{Advanced Audio Coding}
\acrodef{AGU}{Address Generating Unit}
\acrodef{ALU}{Algorithmic Logical Unit}
\acrodef{AVM}{Astute Virtual Machine}
\acrodef{ASIC}{Application-Specific Integrated Circuit}
\acrodef{BSIM}{Berkeley Short-channel IGFET Model}
\acrodef{BTBT}{band-to-band tunneling}
\acrodef{CISC}{Complex Instruction Set Computer}
\acrodef{CMOS}{complementary metal-oxide-semiconductor}
\acrodef{CPU}{Central Processing Unit}
\acrodef{DFG}{Data Flow Graph}
\acrodef{DDIO}{Data Direct I/O}
\acrodef{DLP}{Data Level Parallelism}
\acrodef{DMA}{Direct Memory Access}
\acrodef{DRAM}{Dynamic Random-Access Memory}
\acrodef{DSP}{Digital Signal Processor}
\acrodef{DVS}{Dynamic Voltage Scaling}
\acrodef{DVFS}{Dynamic Voltage and Frequency Scaling}
\acrodef{DPM}{Dynamic Power Management}
\acrodef{EEPROM}{Electrically Erasable Programmable Read-Only Memory}
\acrodef{EER}{Energy Efficiency Rating}
\acrodef{FFT}{Fast Fourier Transformation}
\acrodef{FPA}{floating Point Adder}
\acrodef{FPU}{Floating Point Unit}
\acrodef{FPM}{Floating Point Multiplier}
\acrodef{FMA}{fused multiplyÐadd}
\acrodef{FSM}{Finite State Machine}
\acrodef{GPU}{Graphics Processing Unit}
\acrodef{GPS}{Global Positioning System}
\acrodef{GSM}{Global System for Mobile Communications}
\acrodef{HC}{Hardware Counter}
\acrodef{HDL}{Hardware Description Language}
\acrodef{HPC}{High Performance Computing}
\acrodef{IC}{Integrated Circuit}
\acrodef{IQR}{interquartile range}
\acrodef{ITRS}{Inter\-national Technology Road\-map for Semi\-con\-duc\-tors}
\acrodef{ILP}{Instruction Level Parallelism}
\acrodef{I/O}{input/output}
\acrodef{ISA}{Instruction Set Architecture}
\acrodef{JIT}{Just-In-Time}
\acrodef{JNI}{Java Native Interface}
\acrodef{LCD}{liquid crystal display}
\acrodef{MIPJ}{millions-of-instructions-per-joule}
\acrodef{MOSFET}{metal-oxide semiconductor field-effect transistor}
\acrodef{NIC}{Network Interface Card}
\acrodef{NIST}{National Institute of Standards and Technology}
\acrodef{NDK}{Native Development Kit}
\acrodef{OS}{Operating System}
\acrodef{PID}{proportional-integral-derivative}
\acrodef{RAM}{Random Access Memory}
\acrodef{RISC}{Reduced Instruction Set Computing}
\acrodef{ROHC}{Robust Header Compression}
\acrodef{RMS}{Root Mean Square}
\acrodef{RTL}{Register Transfer Language}
\acrodef{SIMD}{Single Instruction Multiple Data}
\acrodef{SoC}{Systems-on-Chip}
\acrodef{SGLP}{Super-graph Level Parallelism}
\acrodef{SLP}{Super-word Level Parallelism}
\acrodef{SPM}{Scratch-Pad Memory}
\acrodef{SVM}{State Vector Machine}
\acrodef{SRAM}{Static Random-access Memory}
\acrodef{STP}{standard temperature and pressure}
\acrodef{TCP}{Transport Control Protocol}
\acrodef{TCT}{Task Completion Time}
\acrodef{TLB}{Translation Look-aside Buffer}
\acrodef{TLP}{Thread Level Parallelism}
\acrodef{TP}{Travaux Pratiques}
\acrodef{TMU}{Thermal Management Unit}
\acrodef{TTA}{Transport-Triggered Architecture}
\acrodef{UMTS}{Universal Mobile Telecommunications System}
\acrodef{VC}{Virtual Channel}
\acrodef{VM}{Virtual Machine}
\acrodef{VLSI}{Very-Large-Scale Integration}
\acrodef{VHDL}{VHSIC Hardware Description Language}
\acrodef{VLIW}{Very Long Instruction Word}
\acrodef{VM}{Virtual Machine}
\acrodef{WFL}{Weber-Fechner law}
\acrodef{WiFi}{Wireless-Fidelity}
\acrodef{WLAN}{Wireless Local Area Network}
\acrodef{WSN}{Wireless Sensor Network}

\begin{document}

\mainmatter  

\title{The Energy/Frequency Convexity Rule:  Modeling and Experimental Validation on Mobile Devices}

\titlerunning{The Energy/Frequency Convexity Rule}

%
%
\author{Karel De\,Vogeleer\inst{1} \and Gerard Memmi\inst{1} \and Pierre Jouvelot\inst{2} \and Fabien Coelho\inst{2}}
\authorrunning{K. De\,Vogeleer, G. Memmi, P. Jouvelot, and F. Coelho}

\institute{TELECOM ParisTech -- INFRES -- CNRS LTCI - UMR 5141 -- Paris, France \and MINES ParisTech -- CRI -- Fontainebleau, France\\
\mailsa\\
\mailsb}

%
%

\toctitle{Lecture Notes in Computer Science}
\tocauthor{Authors' Instructions}
\maketitle

\begin{abstract}

This paper provides both theoretical and experimental evidence for the existence of an Energy/Frequency Convexity Rule, which relates energy consumption and CPU frequency on mobile devices.
We monitored a typical smartphone running a specific computing-intensive kernel of multiple nested loops written in C using a high-resolution power gauge.
Data gathered during a week-long acquisition campaign suggest that energy consumed per input element is strongly correlated with CPU frequency, and, more interestingly, the curve exhibits a clear minimum over a 0.2\,GHz to 1.6\,GHz window.
We provide and motivate an analytical model for this behavior, which fits well with the data.
Our work should be of clear interest to researchers focusing on energy usage and minimization for mobile devices, and provide new insights for optimization opportunities. 

\end{abstract}

\section{Introduction}
The service uptime of battery-powered devices, e.g., smartphones, is a sensitive issue for nearly any user \cite{6178833}.
Even though battery capacity and performance are hoped to increase steadily over time, improving the energy efficiency of current battery-powered systems is essential because users expect right now communication devices to provide data access every time, everywhere to everyone.
Understanding the energy consumption of the different features of (battery-powered) computer systems is thus a key issue. Providing models for energy consumption can pave the way to energy optimization, by design and at run time.

The power consumption of \acp{CPU} and external memory systems is application and user behavior dependent~\cite{Carroll:2010:APC:1855840.1855861}. Moreover, for cache-intensive and \ac{CPU}-bound applications, or for specific \ac{DVFS} settings, the \ac{CPU} energy consumption may dominate the external memory consumption \cite{Snowdon_RH_05}.
For example, Aaron and Carroll~\cite{Carroll:2010:APC:1855840.1855861} showed that, for an embedded system running \texttt{equake}, \texttt{vpr}, and \texttt{gzip} from the {\small SPEC CPU2000} benchmark suite, the \ac{CPU} energy consumption exceeds the {\small RAM} memory consumption, whereas \texttt{crafty} and \texttt{mcf} from the same suite showed to be straining more energy from the device {\small RAM} memory.

Providing an accurate model of energy consumption for embedded and, more generally, energy-limited devices such as mobile phones is of key import to both users and system designers.
To reach that goal, our paper provides both theoretical and first experimental evidence for the existence of an Energy/Frequency Convexity Rule, that relates energy consumption and CPU frequency on mobile devices.
This convexity property seems to ensure the existence of an optimal frequency where energy usage is minimal. 

This existence claim is based on both theoretical and practical evidence.
More specifically, we monitored a Samsung Galaxy SII smartphone running Gold-Rader's Bit Reverse algorithm~\cite{nla.cat-vn2510360}, a small kernel based on multiple nested loops written in C, with a high-resolution power gauge from Monsoon Solutions Inc.
Data gathered during a week-long acquisition campaign suggest that energy consumed per input element is strongly correlated with CPU frequency and, more interestingly, that the corresponding curve exhibits a clear minimum over a 0.2\,GHz to 1.6\,GHz window.
We also provide and motivate an analytical model of this behavior, which fits well with the data.
Our work should be of clear interest to researchers focusing on energy usage and minimization on mobile devices, and provide new insights for optimization opportunities.

The paper is organized as follows. Section \ref{sec:back} introduces the notions of energy and power, and how these can be decomposed in different components on electronic devices.
Section \ref{sec:testbed} describes the power measurement protocol and methodology driving our experiments, and the {\smaller C} benchmark we used.
Section \ref{sec:model} introduces our \ac{CPU} energy consumption model, and shows that it fits well with the data. Section \ref{sec:convexity} outlines the Energy/Frequency Rule derived from our experiment and modeling. Related work is surveyed in Section \ref{sec:relatedwork}.
We conclude and discuss future work in Section \ref{sec:conclusion}.


\section{Power Usage in Computer Systems}
\label{sec:back}
The total power $\Pt$ consumed by a computer system, including a \ac{CPU}, may be separated into two components: $ \Pt = \Ps + \Pcpu$, where $\Pcpu$ is consumed by the \ac{CPU} itself and $\Ps$ by the rest of system.
In a battery-powered hand-held computer device $\Ps$ may include the power needed to light the {\small LCD} display, to enable and maintain {I/O} devices (including memory), to keep sensors online (\acs{GPS}, gyro-sensors etc.), and others.

The power consumption $\Pcpu$ of the \ac{CPU} we focus on here can be divided into two parts: $\Pcpu = \Pd + \Pl$, 
where $\Pd$ is the power consumed by the \ac{CPU} during the switching activities of transistors during computation.
$\Pl$ is power originating from leakage effects inherent to silicon-based transistors, and is in essence not useful for the \ac{CPU}'s purposes.
$\Pd$ may be split into the power $\Psc$ lost when transistors briefly \emph{short-circuit} during gate state changes and $\Pc$, needed to charge the gates' capacitors:
$ \Pd = \Psc + \Pc.$\label{eq:ptotal}
In the literature $\Pc$ is usually~\cite{Weste:1985:PCV:3928} defined as
$\alpha \,Cf V^2
$, where $\alpha$ is a proportional constant indicating the percentage of the system that is active or switching, $C$ the capacitance of the system, $f$ the frequency at which the system is switching and $V$ the voltage swing across $C$.



$\Psc$ originates during the toggling of a logic gate.
During this switching, the transistors inside the gate may conduct simultaneously for a very short time, creating a direct path between $\Vcc$ and the ground.
Even though this peak current happens over a very small time interval, given current high clock frequencies and large amount of logic gates, the short-circuit current may be non-negligible.
Quantifying $\Psc$ is gate specific but it may be approximated by deeming it proportional to $\Pc$.
Thus the power $\Pd$ stemming from the switching activities and the short-circuit currents in a \ac{CPU} is thus  $\Pc + (\eta-1) \Pc \nonumber$, i.e., $\eta \cdot \alpha C_L f V^2$,
where $\eta$ is a scaling factor representing the effects of short-circuit power.

$\Pl$ originates from leakage currents that flow between differently doped parts of a \ac{MOSFET}, the basic building block of \acp{CPU}.
The energy in these currents are lost and do not contribute to the information that is held by the transistor.
Some leakage currents are induced during the \emph{on} or \emph{off}-state of the transistor, or both.
Six distinct sources of leakage are identified~\cite{Liu:M00/48}.
Despite the presence of multiple sources of leakage in \ac{MOSFET} transistors, the sub-threshold leakage current, gate leakage, and \ac{BTBT} dominate the others for sub-100\,nm technologies~\cite{1468683}.
Leakage current models, e.g., as incorporated in the \BSIM\cite{Liu:M00/48} micro models, are accurate yet complex since they depend on multiple variables.
Moreover, $\Pl$ fluctuates constantly as it also depends on the temperature of the system.
Consequently $\Pl$ cannot be considered a static part of the system's power consumption.
Given the different sources of power consumption in a \ac{MOSFET} based \ac{CPU}, the potal power can be rewritten as $\Pt = \Ps + \Pl + \Pd$. 

The relationship between the \emph{power} $P(t)$ (Watts or Joules/s) and the \emph{energy} $E(\Delta t)$ (Joules) consumed by an electrical system over a time period $\Delta t$ is given by
\begin{equation}
 E(\Delta t) = \int_0^{\Delta t} P(t)~\mathit{dt} = \int_0^{\Delta t} I(t)\cdot V(t)~\mathit{dt}, \label{eq:power}
\end{equation}
where $I(t)$ is the current supplied to the system, and $V(t)$ the voltage drop over the system.
Often $V(t)$ is constant over time, hence $dP(t)/dt$ only depends on $I(t)$.
If both \emph{current} and \emph{voltage} are constant over time, the energy integral becomes the product of \emph{voltage}, \emph{current} and \emph{time}, or alternatively \emph{power} and \emph{time}.

\section{Power Measurement Protocol on Mobile Devices}
\label{sec:testbed}

A Samsung Galaxy S2 is used in our testbed sporting the Samsung Exynos 4 \ac{SoC} 45\,nm dual-core.
The Galaxy S2 has a 32\,KB L1 data and instruction cache, and a 1\,MB L2 cache. 
The mobile device runs Android 4.0.3 on the Siyah kernel adopting Linux 3.0.31.
The frequency scaling governor in Linux was set to operate in \emph{userspace} mode to prevent frequency and voltage scaling on-the-fly.
The second \ac{CPU} core was disabled during measurements.
The smartphone is booted in \emph{clockwork recovery} mode to minimize noisy side-effects of the \ac{OS} and other frameworks.

During the experiments, the phone's battery was replaced by a power supply (Monsoon Power Monitor) that measures the power consumption at 5\,kHz with an accuracy of 1\,mW. The power of the system and the temperature of the \ac{CPU} were simultaneously logged.
The kernel was patched to print a temperature sample to the kernel debug output at a rate of 2\,Hz.

The \emph{bit-reverse} algorithm is used as benchmark kernel.
This is an important operation since it is part of the ubiquitous \ac{FFT} algorithm, and rearranges deterministically elements in an array.
The bit-reversal kernel is \ac{CPU} intensive, induces cache effects, and is economically pertinent.
The Gold-Rader implementation of the bit-reverse algorithm, often considered the reference implementation~\cite{nla.cat-vn2510360}, is given below:

{\small \begin{verbatim}
void bitreverse_gold_rader (int N, complex *data) {
    int n = N, nm1 = n-1; int i = 0, j = 0;
    for (; i < nm1; i++) {
        int k = n >> 1;
        if (i < j) {
            complex temp = data[i]; data[i] = data[j]; data[j] = temp;
        }
        while (k <= j) {j -= k; k >>= 1;}
        j += k;
    }
}
\end{verbatim}
}

The input of the bit-reversal algorithm is an array with a size of $2^N$; the elements are pairs of 32\,bit integers, representing complex numbers. Note that array sizes up to $2^{9}$ fit in the L1 cache, while sizes over $2^{18}$ are too big to fit in the L2 cache.

During the measurements, $N$ is set between 6 and 20 in steps of 2, while varying the \ac{CPU} frequency from 0.2\,GHz to 1.6\,GHz in steps of 0.1.

To minimize overhead, 128 copies of the kernel are run sequentially. For time measurement purposes, this benchmark is repeated 32 times for at least 3 seconds each time (this may require multiple runs of the 128 copies).
For the power and temperature measurements, the benchmark is repeated in an infinite loop until 32 samples can be gathered.
The benchmark is compiled with {\small GCC} 4.6, included in Google's {\small NDK}, generating {\small ARM}v5 thumb code.

Data was fitted using \texttt{R} and the \texttt{nls()} function employing the \emph{port} algorithm.


\section{Modeling Energy Consumption}
\label{sec:model}

Energy is the product of time by power. We look at each of these factors in turn here.

\subsection{Execution Time}
\label{sec:time}

Since applications run over an \ac{OS}, we need to take account for it when estimating computing time. Indeed, an \ac{OS} needs a specific amount of time, or clock cycles, to perform (periodical) tasks, e.g., interrupt handling, process scheduling, processing kernel events, managing memory etc.
When the processor is not spending time in kernel mode, the processor is available for user-space programs, e.g., our benchmark.
From a heuristic point of view, it can be assumed that the \ac{OS} kernel needs a fixed amount of clock cycles $\ck$ per time unit to complete its tasks.
Thus, we propose to model the amount of clock cycles to complete a benchmark sequence of instructions $\cb$ as $t (f^\beta - \ck)$,
where $\ck$ are the number of clock cycles spent in the \ac{OS}, $t$ the total time needed to complete the program, $f$ the system's clock frequency and $\beta$ an architecture-dependent scaling constant, to be fitted later on with the data.
The definition of $\cb$ is rewritten to isolate the execution time:
\begin{equation}
 t = \frac{\cb}{f^\beta - \ck}.\label{eq:time}
\end{equation} 
Note that $t$ tends to zero for $f\rightarrow\infty$ and there is a vertical asymptote at $\sqrt[\beta]{\ck}$.


Table \ref{table:timeerror} shows the fitting errors of Equation \ref{eq:time} on the execution time measurement data, averaged over all tested input sizes.
\begin{table}[b]
\begin{center}
\caption{Average absolute execution time ($t$), power ($P$), and energy ($E$) fitting errors (\%) of Equation \ref{eq:time}, Equation  \ref{eq:ptfit}, and Equation  \ref{eq:energy} respectively, on the measured data given different CPU frequencies (f) at a 37$^\circ$C core temperature.\label{table:timeerror}}
\setlength{\tabcolsep}{2pt}
\begin{tabular}{l|ccccccccccccccc}
   {$\mathbf{f}$ (GHz)} & 0.2 & 0.3 & 0.4 & 0.5 & 0.6 & 0.7 & 0.8 & 0.9 & 1.0 & 1.1 & 1.2 & 1.3 & 1.4 & 1.5 & 1.6 \\\hline
   {$\mathbf{error~}t$} & 1.18 & 2.71 & 1.55 & 0.55 & 0.56 & 4.21 & 0.62 & 1.63 & 4.71 & 3.68 & 1.86 & 0.44 & 5.87 & 0.75 & 2.61 \\
   {$\mathbf{error~}P$} &  2.94 & 1.00 & 0.20 & 0.99 & 1.31 & 1.46 & 1.24 & 0.49 & 0.02 & 0.70 & 0.86 & 0.82 & 0.03 & 7.40 & 0.58 \\
   {$\mathbf{error~}E$} & 18.39 & 0.83 & 0.92 & 2.93 & 3.31 & 1.34 & 2.73 & 2.37 & 4.69 & 4.68 & 3.01 & 1.46 & 5.83 & 8.02 & 3.27 \\
\end{tabular} 
\end{center}
\end{table} 
The fitting exhibits a vertical asymptote around 115\,MHz.
This may indicate the minimum amount of clock cycles required by the \ac{OS} of the phone to operate.
The measurement data for input sizes $2^6$ up to $2^{16}$ are well described by Equation~\ref{eq:time}.
However, sizes $2^{18}$ and $2^{20}$, too large to fit within cache L2, seem to operate under different laws.
Therefore, from now on, the attention is focused on data that fit in the cache of the \ac{CPU}.


\subsection{Power Consumption}
\label{sec:leakage}


If dynamic power modeling is rather easy (see Section \ref{sec:back}), the case for leakage is more involved, and warrant a longer presentation. In particular, leakage power is heavily temperature-dependent~\cite{Liu:M00/48}.
For example, our \ac{CPU} at 1.3\,GHz shows an inflated power consumption of around 5\% between a \ac{CPU} temperature of 36$^{\circ}$C and 46$^{\circ}$C.
You~\ea\cite{You:2002:CAS:2144385.2144389} shows similar results for a 0.1$\mu$m processor; a temperature increase from 30$^\circ$C to 40$^\circ$C leads to a 3\% power increase.
On the other hand, the power $\Pc$ required for a given computation does not change with regards to the \ac{CPU} temperature.
The Berkeley Short-channel IGFET Model ({\smaller BSIM}) \cite{Liu:M00/48} shows that the leakage current micro models depend on a multitude of variables.
The temperature itself appears several times in the sub-threshold and \ac{BTBT} leakage models; the gate leakage however is not temperature dependent.
Mukhopadhyay \ea\cite{1218931} showed via simulation that for 25\,nm technology the sub-threshold leakage current is dominant over the \ac{BTBT} leakage current, but the latter cannot be neglected.
Under normal conditions, the temperature of the \ac{CPU}'s silicon varies continuously depending on the load of the \ac{CPU} and the system's ambient temperature.
Therefore, to have a fair comparison of energy consumption between different code pieces one needs to compare the measurements at a reference temperature.

Finding a temperature scaling factor for the leakage current is however not a straightforward task.
Nevertheless, approximative scaling factors have been analytically obtained or experimentally defined via simulations (mainly {\small SPICE})~\cite{Su:2003:FCL:871506.871529,Liao:2006:TSV:2298535.2301211,814859}.
After analysis on our data, we discovered that none of the cited approximations would fit well.
This is because the rationale on which these approximations are based assume conditions, which are not entirely realistic, to simplify the leakage current micro models.
Most previous research works focus solely on the sub-threshold leakage effect, neglecting other leakage effects.
This may be appropriate for technologies larger than the 45\,nm technology we use.

Skadron \ea\cite{Skadron:2004:TMM:980152.980157} studied the temperature dependence of $\Il$ as well.
Skadron \ea deducted a relationship between the leakage power $\Pl$ and dynamic power $\Pd$ based on \acf{ITRS} measurement traces (variables indexed with 0 are reference values):
\begin{equation}
  R_T = \frac{\Pl}{\Pd} = \frac{R_0}{V_0T^2_0} e^{\frac{B}{T_0}}  V T^2  e^{\frac{-B}{T}}.\label{eq:Rt}
\end{equation}
If the temperature $T$ is stable across different operating voltages, then the value of $R_T$ is a function of $V$ multiplied by a constant $\gamma$, which includes the temperature dependent variables and other constants. Total power $\Pt$ is thus:
\begin{eqnarray}
\Pt  & = & \Ps + \Pl + \Pd \nonumber\\
     & = & \Ps + \gamma V\Pd + \Pd\nonumber\\
     & = & \Ps + (1+\gamma V)\cdot \eta\alpha C f V^2.\label{eq:ptfit}
\end{eqnarray}
This formulation of $\Pt$ incorporates three parameters: $\Ps$, $\gamma$, and $\eta\alpha C$.
The values of these variables can be obtained via fitting power traces on Equation~\ref{eq:ptfit}.
$V$ and $f$ are linked via the \ac{DVFS} process inherent to the Linux kernel and the hardware technicalities.
Experimental values for our \ac{CPU} are found inside the Siyah kernel; they are shown in Table~\ref{table:dfvs}.
\begin{table}[t]
\begin{center}
\caption{Frequency and voltage relationship for the \acl{DVFS} (DVFS) process in the default Siyah kernel.\label{table:dfvs}}
\begin{tabular}{c|ccccccccccccccc}
   {$\mathbf{f}$ (MHz)}\, & ~200 & 300 & 400 & 500 & 600 & 700 & 800 & 900 & 1000 & 1100 & 1200 & 1300 & 1400 & 1500 & 1600 \\\hline
   {$\mathbf{V}$ (mV)}\, & ~920 & 950 & 950 & 950 & 975 & 1000 & 1025 & 1075 & 1125 & 1175 & 1225 & 1250 & 1275 & 1325 & 1350 \\
\end{tabular} 
\end{center}
\end{table} 
The power fitting errors are shown in Table~\ref{table:timeerror} for a 37$^\circ$C \ac{CPU} temperature.
The fitting errors are on the average not larger than 3\,\% except for the measurement point at 1.5\,GHz.
This measurement point was obtained at different independent occasions but appears, for obscure reasons, to disobey persistently the model in Equation \ref{eq:ptfit}.




\subsection{Energy Consumption}
\label{sec:energy}

Typical compute-intensive programs incur approximately a constant load on the \ac{CPU} and system, barring user interactions.
Moreover, if the time to complete one program is also much smaller than the sampling rate of the power gauge, then $P(t)$ in Equation~\ref{eq:power} is constant.
Hence, it suffices to sample the power $\Pb$ of a benchmark at a given \ac{CPU} temperature and multiply this value by the execution time of the benchmark $\tb$ to get an energy estimate.
As a result the definition of time in Equation~\ref{eq:time}, and power in Equation~\ref{eq:ptfit}, can be used to model the energy of one benchmark kernel.
The energy consumed by the \ac{CPU} $\Ecpu$ is given by
\begin{eqnarray}
\Ecpu           & = & \El + \Ed \nonumber\\
                & = & \Pb \cdot \tb \nonumber \\
                & = & \left((1+\gamma V)\cdot \eta \alpha C f V^2\right)\cdot \frac{\cb}{f^\beta - \ck}.\label{eq:energy}
\end{eqnarray}
Constants $\gamma$, $\eta\alpha C$, $\cb$, and $\ck$ in this formulation were evaluated before via fitting the power and time traces.

\section{The Energy/Frequency Convexity Rule}
\label{sec:convexity}

\begin{figure}[t]
\label{fig:Ebfigure}
\centering
        \input{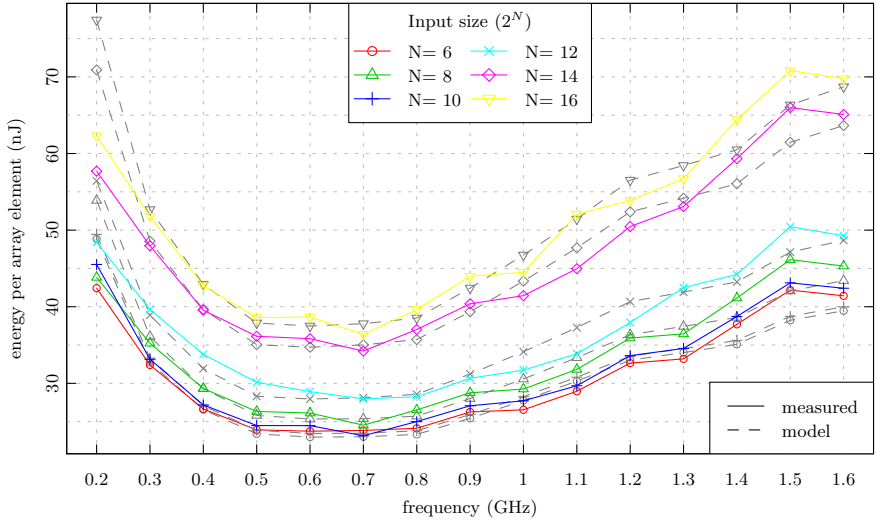}
\caption{Energy required by the CPU at 37$^\circ$C to complete our benchmark kernel given an input size.
          The dashed lines denote the theoretical curve as per Equation~\ref{eq:energy}.}
\end{figure}

Using the testbed and models described above, Figure~\ref{fig:Ebfigure} shows the measured and modeled energy $\Ecpu$ for our benchmark kernel over the different frequencies; data have been normalized over the benchmark input size.
Table~\ref{table:timeerror} shows the average absolute energy error between our fitted model and the measured data.
The average fitting error stays below 6\,\% except for measurement points 1.5\,GHz and 200\,MHz.
The large fitting error in the 200\,MHz case stems from the large execution time that amplifies the power measurement fitting error (see Table~\ref{table:timeerror}).
It can also be seen that, for larger benchmark input sizes, on the average more energy is required.
This is the result of higher level cache utilization.

Figure~\ref{fig:Ebfigure} exhibits a clear convex curve, with a minimum at Frequency $\fo$, suggesting the existence of an Energy/Frequency Convexity Rule for compute-intensive programs. Why is the energy consumption curve convex?
The energy consumption of the benchmark kernel scales approximately linearly with the number of instructions.
The time $\Delta t$ it takes to execute an instruction sequence increases more than linearly with decreasing operating frequency.
$\Pl$ is independent of the type of computation, and $\El$ builds up linearly with time: $\El = \Pl \Delta t$.
$\Pl$ becomes increasingly important in the part where the \ac{CPU} frequency $f$ is smaller than $\fo$.
For the part where $f>\fo$, the inflated $\Ecpu$ can be attributed to the increasing supply voltage $\Vcc$ affecting $\Pd$ quadratically.
Furthermore, had $\Ps$ been incorporated in the picture, then $\fo$ would have moved to a higher frequency because the additional consumed energy of the system could have been minimized by a faster completion of the computations on the \ac{CPU}.

Our proposal for the existence of an Energy/Frequency Convexity Rule can be further supported using our previous models. Indeed, we can model the relationship in Table \ref{table:dfvs} between the \emph{frequency} (GHz) and \emph{voltage} (V) in the Linux kernel with a linear approximation:
$ V = m_1 f+ m_2\label{eq:Vfrel}.$ Now the derivative of $\Ecpu$ defined in Equation~\ref{eq:energy} over $f$ or $V$ can be computed.
The energy curve shows a global minimum $\Ecpum$ for $\fo$ when its derivative is equal to zero ($\partial\Ecpu/\partial f=0$) and its second derivative is positive.


Given that $\Ecpu$ only shows one minimum, $\fo$ is the global minimum if the following equality holds:
\begin{equation}
\frac{(1+\gamma V) V f^\beta \beta}{f^\beta - \ck} = f m_1 (3 \gamma V + 2 ) + (1+\gamma V) V. \label{eq:derivEb}
\end{equation}
Four parameters appear in this formulation that affect the optimal frequency $\fo$: $\beta$ and $\ck$, which are related to the execution time of the benchmark, $m_1$ the slope between $V$ and $f$, and $\gamma$ related to the leakage current ratio. Simulations show that, if $\beta$ or $\ck$ decreases, $\Ecpum$ will shift to a higher frequency, and, if $m_1$ or $\gamma$ decreases, $\fo$ will decrease as well.
$\gamma$ is temperature dependent; if the temperature increases, $\gamma$ will increase accordingly.
Hence, $\Ecpum$ and $\fo$ increase with temperature as well.
For the presented measurements Equation~\ref{eq:energy} shows a minimum on the average around 700\,MHz.
This holds for all input sizes of the benchmark between $2^{6}$ and $2^{16}$.
As a result, this implies that there exists an operating frequency, which is neither the maximum nor the minimum operating frequency, at which the \ac{CPU} would execute a code sequence on the top of the OS in the most energy efficient way.

\section{Related Work}
\label{sec:relatedwork}

The convex property of the energy consumption curve has been hinted at before in the literature. A series of papers, approaching the problem from an architectural point of view, have suggested a convex energy consumption curve with respect to DVFS ~\cite{Snowdon_RH_05,Fan:2003:SPM:2157911.2157927,LeSueur:2010:DVF:1924920.1924921}.
The authors put forward some motivation for the convexity of the curve, but do not provide an analytical framework.
A detailed explanation or an analytical derivation, as presented here, is not given however.


In the \ac{VLSI} design domain voltage scaling has also been discussed but usually for a fixed frequency~\cite{Zhai:2004:TPL:996566.996798,4271869}.
The aim of the voltage scaling is to find a minimum energy operation point where the digital circuit yields the correct output.
The major trade-off is between increasing circuit latency and leakage power, and decreasing dynamic power.
This trade-off yields also a convex energy consumption curve, but for a fixed frequency.


\section{Conclusion}
We provide an analytical model to describe the energy consumption of a code sequence running on top of the \ac{OS} of a mobile device. The energy model is parametrized over five parameters abstracting the specifics of the \acf{DVFS} process, the execution time related parameters, and the power specifications of the \ac{CPU}.
Measurement traces from a mobile device were used to validate the appropriately fitted model.
It is shown that the model is on the average more than 6\,\% accurate.
The importance of power samples obtained at a reference temperature is also pointed out.

It is also shown that the analytical energy model is convex (representing what we call the Energy/Frequency Convexity Rule) and yields a minimum energy consumption of a code sequence for a given \ac{CPU} operation frequency. This minimum is a function of the temperature, execution time related parameters, and technical parameters related to the hardware.
A more in depth analysis of the Energy/Frequency Convexity Rule can be found in our technical report \cite{kdv2013}.

Future work includes checking the validity of our model and its parameters over a wide range of compute-intensive benchmarks. Also, extending the presented model to better handle memory access operations, in particular the impact of caches, is deserved.
Finally a generalization of the model to encompass the impact of other programs running in parallel with benchmarks or system power effects would be useful.

\label{sec:conclusion}



\bibliographystyle{IEEEtran}
\bibliography{CP120_library}

\end{document}